\documentclass[prd,10pt,aps,showpacs,onecolumn,longbibliography,nofootinbib,amsmath,amssymb,superscriptaddress, notitlepage]{revtex4-1}

\usepackage[T2A]{fontenc}
\usepackage[utf8]{inputenc}
\usepackage[english]{babel}
\usepackage[a4paper,margin=2.5cm]{geometry}
\usepackage{amsmath,amssymb,amsfonts,amsthm,mathtools}
\usepackage{bm}
\usepackage{mathrsfs}
\usepackage{physics}

\usepackage{graphicx}
\usepackage{float}
\usepackage{caption}
\usepackage{subcaption}

\usepackage{array}
\usepackage{booktabs}
\usepackage{multirow}

\usepackage{enumitem}

\usepackage{hyperref}
\usepackage[nameinlink,noabbrev]{cleveref}

\usepackage{microtype}

\usepackage{comment}

\numberwithin{equation}{section}

\renewcommand{\vec}[1]{\mathbf{#1}}

\begin{document}

\title{Monte Carlo Analysis of Topological Unfolding Pathways in Complex Knotted Proteins}
\author{A. M. Begun}
\affiliation{Pacific Quantum Center, Far Eastern Federal University, 10 Ajax settlement, Russkiy Island, 690022, Vladivostok, Russia}
\author{S. D. Liubimov}
\affiliation{Pacific Quantum Center, Far Eastern Federal University, 10 Ajax settlement, Russkiy Island, 690022, Vladivostok, Russia}
\author{N. V. Gerasimeniuk}
\affiliation{Pacific Quantum Center, Far Eastern Federal University, 10 Ajax settlement, Russkiy Island, 690022, Vladivostok, Russia}
\author{A. V. Zorina}
\affiliation{Pacific Quantum Center, Far Eastern Federal University, 10 Ajax settlement, Russkiy Island, 690022, Vladivostok, Russia}
\author{I. A. Baldin}
\affiliation{Pacific Quantum Center, Far Eastern Federal University, 10 Ajax settlement, Russkiy Island, 690022, Vladivostok, Russia}

\begin{abstract}
We study the thermal unfolding of two proteins containing complex knots: the ($5_2$)-knotted UCH-L1 (3IRT) and the ($6_1$)-knotted DL-2-haloacid dehalogenase (4N2X). Their backbones are described by an effective Landau--Ginzburg--Wilson soliton model, and unfolding is simulated using Glauber Monte Carlo dynamics. The limiting transition temperature ($T_{\infty}$) is estimated by extrapolating simulations with increasing heating duration. The ($5_2$) knot of 3IRT unfolds predominantly through a direct transition to the unknotted state, whereas 4N2X exhibits intermediate ($3_1\#4_1$) and ($4_1$) topologies. The results demonstrate qualitatively different topological unfolding mechanisms for the two complex protein knots.
\end{abstract}

\maketitle
\section{Introduction}

Knots are relatively rare but structurally important features of protein backbones. Although protein chains are open, their folded conformations can contain well-defined knotted regions that can be classified using appropriate topological procedures. Protein knots have been associated with nontrivial folding mechanisms, kinetic barriers, and complex unfolding behavior~\cite{Virnau2011,DabrowskiTumanski2019}. While the trefoil knot \(3_1\) is the simplest and most common nontrivial protein knot, several proteins contain more complex topologies, including the twist knots \(5_2\) and \(6_1\).

The mechanisms by which such complex knots are removed during unfolding remain less well understood. Previous studies of the \(5_2\)-knotted protein UCH-L1 have shown that unknotting can proceed either directly or through a transient \(3_1\) state, depending on the direction of unthreading~\cite{Zhao2018,UCHL1SingleMolecule}. For \(6_1\)-knotted dehalogenases, simulations have suggested pathways involving a simpler \(4_1\) topology, while experiments indicate the presence of multiple conformational intermediates during folding and unfolding~\cite{DehISimulation,DehIFolding}. These observations suggest that the loss of a complex protein knot may involve several distinct topological rearrangements rather than a single transition to the unknotted state.

In the present work, we investigate the topological unfolding of two proteins containing different complex twist knots: the \(5_2\)-knotted UCH-L1 structure 3IRT and the \(6_1\)-knotted DL-2-haloacid dehalogenase 4N2X. Their structures are represented using an effective Landau--Ginzburg--Wilson soliton model, and thermal unfolding is simulated by Glauber Monte Carlo dynamics. By varying the duration of the heating protocol, we determine the limiting transition temperature \(T_{\infty}\) and analyze the sequence of knot types visited during unfolding. 

\section{Materials and methods}

\subsection{Topological knots in polymer and protein chains}

A protein backbone can be represented as an open spatial curve defined by the coordinates of its monomers or \(C_{\alpha}\) atoms. In mathematical knot theory, however, a knot is defined for a closed curve. Therefore, conventional methods for determining the knot type of a protein usually require an artificial closure connecting the N- and C-termini. The resulting classification may depend on the selected closure procedure~\cite{Virnau2011,DabrowskiTumanski2019}.

In the present work, the protein chain was not explicitly closed in three-dimensional space. Instead, its topology was determined using the \texttt{Knot\_pull} algorithm~\cite{Jarmolinska2020}. The method iteratively smooths and simplifies the open chain while keeping its termini fixed and preventing the backbone from passing through itself. After removing all topologically unnecessary points, the knot type is determined from the Dowker--Thistlethwaite code. The connection between the two termini is introduced only implicitly, in such a way that no additional crossings are created.

\subsection{Knot classification}

\begin{figure}[ht]
    \centering
    \includegraphics[width=0.9\textwidth]{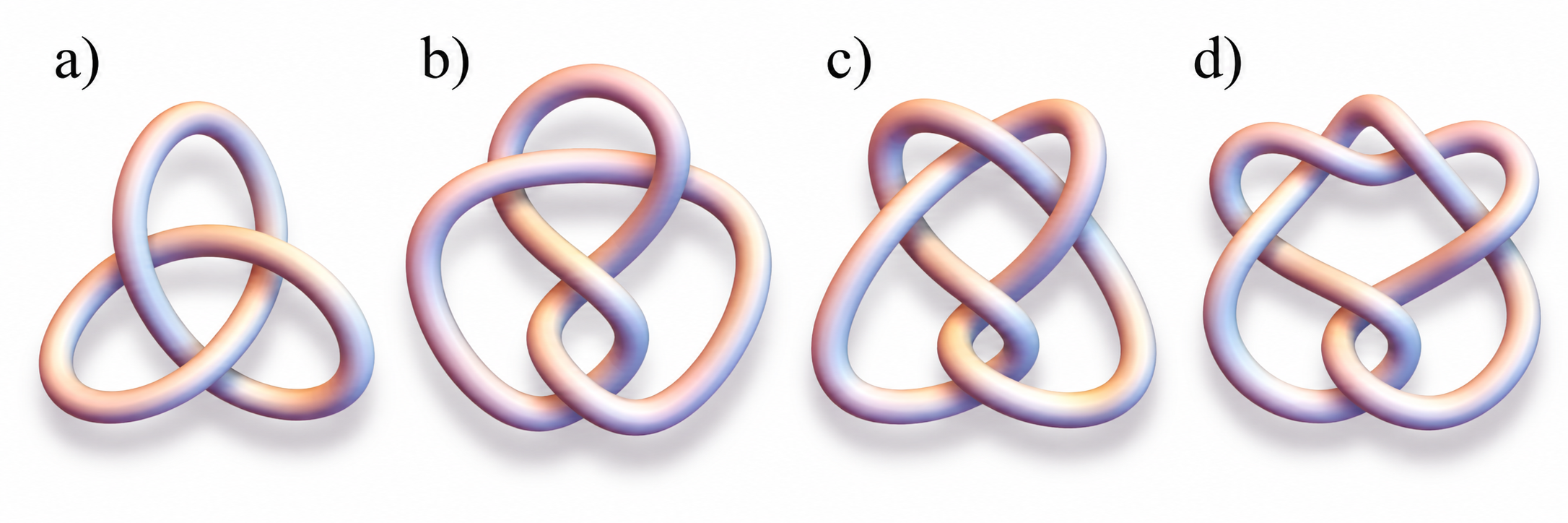}
    \caption{Examples of different knot types: \(3_1\), \(4_1\), \(5_2\), and \(6_1\).}
    \label{fig:knot_types}
\end{figure}

Knot types are denoted using the Alexander--Briggs notation \(n_m\), where \(n\) is the minimal number of crossings in a planar diagram, while the index \(m\) distinguishes different knots with the same crossing number. The simplest nontrivial knot is the trefoil knot \(3_1\). The figure-eight knot \(4_1\) is the only prime knot with four crossings. There are two different prime knots with five crossings: \(5_1\), also known as the cinquefoil knot, and \(5_2\), usually referred to as the three-twist knot. The knot \(6_1\) is one of the prime knots with six crossings and is commonly known as the stevedore knot~\cite{AlexanderBriggs1927,Hoste1998}. The corresponding structures of the knots \(3_1\), \(4_1\), \(5_2\), and \(6_1\) are shown in Fig.~\ref{fig:knot_types}.

The knots \(5_2\) and \(6_1\) belong to the family of twist knots. A twist knot can be constructed by forming a loop from two nearby segments of a curve, introducing a number of successive half-twists between them, and closing the remaining ends by a clasp. Changing the number and orientation of the half-twists produces different members of this family. In particular, the trefoil knot \(3_1\), the figure-eight knot \(4_1\), the three-twist knot \(5_2\), and the stevedore knot \(6_1\) can all be represented as twist knots~\cite{Rolfsen1976}.

In addition to prime knots, composite knots can be constructed by taking connected sums. For example, $
3_1 \mathbin{\#} 4_1$ denotes the connected sum of the trefoil knot \(3_1\) and the figure-eight knot \(4_1\).  The obtained curve is a single closed knot containing the topological structures of both initial knots~\cite{Rolfsen1976}.

In what follows, the topological unfolding pathway is defined as the sequence of topological states adopted by the protein chain during its transition from the native knotted conformation to a fully unfolded or topologically trivial state.

\subsection{Frenet coordinates}

\begin{figure}[ht]
    \centering
    \includegraphics[width=0.8\textwidth]{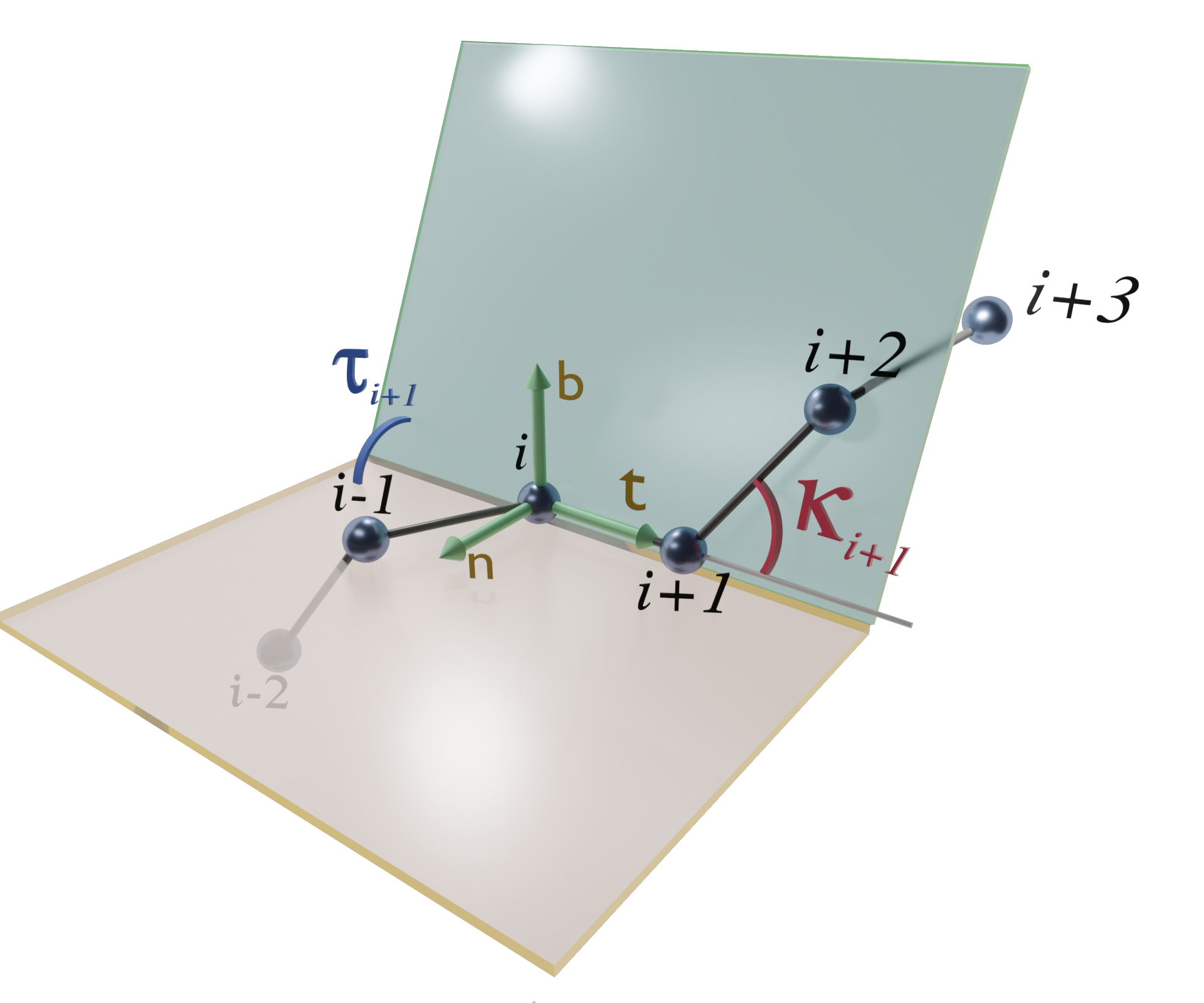}
    \caption{Geometrical representation of a protein backbone as a piecewise linear curve and definition of the discrete Frenet frame. The local frame at site $i$ is formed by the tangent $\mathbf{t}_i$, normal $\mathbf{n}_i$, and binormal $\mathbf{b}_i$ vectors. The curvature angle $\kappa_i$ describes the change in the direction of consecutive tangent vectors, while the torsion angle $\tau_i$ describes the rotation of consecutive binormal vectors.}
    \label{fig:frenet_frame}
\end{figure}

The protein backbone is represented as a piecewise linear curve connecting consecutive $C_\alpha$ atoms. Its geometry is described in terms of discrete Frenet coordinates. At each site of the curve, one can introduce the Frenet frame formed by three mutually orthogonal vectors: the tangent vector $\mathbf{t}_i$, the normal vector $\mathbf{n}_i$, and the binormal vector $\mathbf{b}_i$. These vectors provide a local coordinate system attached to the curve and allow its geometry to be described in terms of the curvature and torsion angles~\cite{Hu2011}. The geometrical construction of the discrete Frenet frame and the corresponding curvature and torsion angles is illustrated in Fig.~\ref{fig:frenet_frame}.

\begin{eqnarray}
\vec t_i&=&\frac{\vec r_{i+1}-\vec r_i}{|\vec r_{i+1}-\vec r_i|},\\
\vec b_i&=&\frac{[\vec t_{i-1},\vec t_{i}]}{|[\vec t_{i-1},\vec t_{i}]|},\\
\vec n_i&=&[\vec b_i, \vec t_i],
\end{eqnarray}

where $\vec r_i$ are coordinate vectors of $C_\alpha$ atoms.

The angles of curvature $\kappa_i$ and torsion $\tau_i$ act as new coordinates. They can be calculated as follows:
\begin{eqnarray}
\kappa_i&=&\arccos(\vec t_{i+1}, \vec t_i),\\
\tau_i &=&\mathrm{sign}([\vec b_{i-1},\vec b_i], \vec t_i)\arccos(\vec b_{i+1}, \vec b_i).
\end{eqnarray}

In the continuous limit, the evolution of the Frenet frame along the curve is governed by the Frenet equations, which can be written in matrix form as
\begin{equation}
\frac{d}{ds}
\begin{pmatrix}
\mathbf{t} \\
\mathbf{n} \\
\mathbf{b}
\end{pmatrix}
=
\begin{pmatrix}
0 & \kappa & 0 \\
-\kappa & 0 & \tau \\
0 & -\tau & 0
\end{pmatrix}
\begin{pmatrix}
\mathbf{t} \\
\mathbf{n} \\
\mathbf{b}
\end{pmatrix}.
\label{eq:frenet}
\end{equation}

The space curve defined by the protein backbone is invariant under local $SO(2)$ rotations of the Frenet frame around the tangent vector $\mathbf{t}_i$. Namely, the normal and binormal vectors transform as
\[
\begin{pmatrix}
\mathbf{n}_i \\
\mathbf{b}_i
\end{pmatrix}
\rightarrow
\begin{pmatrix}
\cos\eta_i & \sin\eta_i \\
-\sin\eta_i & \cos\eta_i
\end{pmatrix}
\begin{pmatrix}
\mathbf{n}_i \\
\mathbf{b}_i
\end{pmatrix},
\label{eq:SO2}
\]
without changing the underlying curve~\cite{Danielsson2010,Hu2011}. 

\subsection{Landau-Ginzburg-Wilson approach}
Using the curvature and torsion variables introduced above, we construct an effective Landau--Ginzburg--Wilson description of the protein backbone~\cite{Chernodub2010,Niemi2014}.

Within the Landau--Ginzburg--Wilson approach, the starting point is the Abelian Higgs Hamiltonian written for a complex scalar field $\psi(s)$~\cite{Chernodub2010,Niemi2014}:
\begin{equation}
H=\partial_s\psi\,(\partial_s\psi)^*+\lambda\left(\psi\psi^*-v^2\right)^2.
\label{eq:H_psi}
\end{equation}

We construct the Landau--Ginzburg--Wilson Hamiltonian using the Hasimoto transformation~\cite{Hasimoto1972,Niemi2014}. This transformation introduces a complex scalar field $\psi(s)$, which is related to the geometry of the curve as
\begin{equation}
    \psi(s)=\kappa(s)\exp\left(i\int\limits_{0}^{s}\bigl(\tau(s')+\alpha(s')\bigr)\,ds'\right),
\label{eq:Hasimoto}
\end{equation}
where $\alpha(s)$ is a site-dependent function representing an external field.

The modulus of the scalar field is equal to the curvature $\kappa(s)$, while its phase is determined by the integral of the torsion $\tau(s)$ and the external field $\alpha(s)$. Thus, the Hasimoto transformation replaces the pair of real functions $\kappa(s)$ and $\tau(s)$, which define the curve geometry, by a single complex scalar field $\psi(s)$.

The local rotation of the Frenet frame \eqref{eq:SO2} changes the gauge-dependent variables, while the physical curve remains unchanged. Therefore, the scalar field $\psi_i$, understood as a gauge-invariant order parameter describing the backbone geometry, remains invariant under such $SO(2)\simeq U(1)$ transformations.

While the curvature and torsion variables acquire a gauge-dependent form. In particular, the curvature can be combined with the local rotation angle into a complex quantity,
\[
\kappa(s) \rightarrow \kappa(s)e^{i\eta(s)},
\]
while the torsion transforms as an Abelian gauge field,
\[
\tau(s) \rightarrow \tau(s)-\frac{d\eta(s)}{ds}.
\]

Consequently, the Hasimoto field \eqref{eq:Hasimoto} is invariant under this local transformation. Therefore, all physical quantities constructed from $\psi(s)$ remain invariant as well, including the Hamiltonian \eqref{eq:H_psi}~\cite{Danielsson2010,Niemi2014}.

It is important to note that, in the present model, the function $\alpha(s)$ is not a dynamical field. Instead, it is treated as an external parameter that effectively accounts for the influence of the surrounding medium on the protein chain.

Rewriting the Hamiltonian \eqref{eq:H_psi} in terms of the geometric variables \(\kappa\) and \(\tau\), we obtain
\begin{equation}
H=(\partial_s \kappa)^2+\kappa^2(\tau+\alpha)^2+\lambda\left(\kappa^2-v^2\right)^2.
\label{eq:Hkappatau}
\end{equation}
The first term in \eqref{eq:Hkappatau} is the kinetic contribution associated with spatial inhomogeneities of the curvature. The second term describes the contribution of the local torsion to the energy of the system, while the function \(\alpha(s)\) introduces an additional external contribution to the phase structure of the field \(\psi\). Finally, the last term represents a nonlinear double-well potential that leads to the existence of several vacuum states and therefore, provides the conditions for the emergence of soliton configurations.

Variation of hamiltonian leads to equations of motion:

\begin{eqnarray}
\partial_s^2 \kappa-2\lambda\left(\kappa^2-v^2\right)\kappa&=&0,\\
    \tau+\alpha&=&0,
\end{eqnarray}
and correspnding solutions in the form of kink:
\begin{eqnarray}
\kappa(s)&=&v\,\tanh\bigl(v\sqrt\lambda(s-s_0)\bigr),\\
\label{eq:kink}
\tau&=&-\alpha.
\end{eqnarray}

In the context of protein structure, such kinks are interpreted as local transitions between different stable conformations. Regular secondary structures, such as $\alpha$-helices and $\beta$-strands, correspond to the vacuum states of the effective potential, whereas loops and transition regions are described by soliton configurations connecting different energy minima~\cite{Chernodub2010,Niemi2014}.

For a more realistic description, we introduce an asymmetric potential
\begin{equation}
    V(\kappa)=\lambda(\kappa+v_1)^2(\kappa-v_2)^2,
    \label{eq:asim_pot}
\end{equation}
which corresponds to two non-equivalent vacuum states. The equation of motion for $\kappa$ then takes the form
\begin{equation}
    0=\partial_s^2\kappa-\lambda(\kappa+v_1)(\kappa-v_2)(2\kappa+v_1-v_2).
\end{equation}
As a result, the soliton solutions become asymmetric as well. This generalization is important for describing real proteins, since different structural elements may possess different energetic stability.

As far as protein is discrete partly linear curve connecting aminoacid's residues it is necessary to discretize all the equations. The derivative in the hamiltonian must be replaced by the difference:
\begin{equation}
    \partial_s \psi \rightarrow \frac{1}{a}(\psi_{i+1}-\psi_i),
\label{eq:discrete_derivative}
\end{equation}
As a result, the discrete Hamiltonian takes the form
\begin{equation}
H=\sum_{i=1}^{L-1}\left(
\kappa_{i+1}^2+\kappa_i^2
-2\kappa_{i+1}\kappa_i\cos(\tau_{i+1}+\alpha_{i+1})
+\lambda(\kappa_i^2-v^2)^2
\right),
\label{eq:discrete_H}
\end{equation}
while in the asymmetric case it is written as
\begin{equation}
H=\sum_{i=1}^{L-1}\left(
\kappa_{i+1}^2+\kappa_i^2
-2\kappa_{i+1}\kappa_i\cos(\tau_{i+1}+\alpha_{i+1})
+\lambda(\kappa_i+v_1)^2(\kappa_i-v_2)^2
\right).
\end{equation}
Thus, we obtain a lattice gauge model in which the interaction between neighboring sites is determined by a phase factor containing the torsion and the compensating gauge function~\cite{Danielsson2010,Chernodub2010}.

The resulting structure is built as a sequence of solitons. Each soliton obtain it's own values of $v_1,v_2,\lambda$. $\alpha_i$ is independently defined in each site. The values of all parameters are calculated in such a way to minimize RMSD from experimental structure in the minimum of the Hamiltonian. Simulating annealing algorithm is used to find corresponding values.

\subsection{Unfolding modeling}

Unfolding was modeled using Monte-Carlo methods, the Glauber algorithm in particular~\cite{Glauber1963,Peng2016}. It differs from common Metropolis algorithm in accepting probability.

At each Monte Carlo step, either the curvature or the torsion angle
at a randomly selected site $i$ is perturbed by a small random value:

\begin{equation}
    \kappa_i' = \kappa_i + \sigma_\kappa r,
\end{equation}
\begin{equation}
    \tau_i' = \tau_i + \sigma_\tau r,
\end{equation}
where $r$ is a normally distributed random variable with zero
mean and unit variance, and $\sigma_\kappa$ and $\sigma_\tau$
are the amplitudes of the corresponding perturbations.

To preserve the topology of the continuous protein backbone,
self-crossings must be prevented during the simulation.
For this purpose, we impose a self-avoidance condition requiring
the minimum distance between non-neighboring $C_\alpha$ atoms
to exceed $3.8\,\text{\AA}$.
If this condition is violated, the proposed configuration is rejected.
Otherwise, the acceptance probability is calculated as follows:

\begin{equation}
    P=\frac{1}{1+e^{\Delta H/T_G}}.
\end{equation}

The unfolding was modeled as quasistatic heating of the protein. 
During heating temperature $T_G$ is increasing by exponential law with growth factor defined by required amount of steps and temperature range:
\begin{equation}
  T_G'=T_Gb,  
\end{equation}
\begin{equation}
  b=\exp{\ln(\frac{T_{max}}{T_{min}})N^{-1} },
\end{equation}
where $b$ is the temperature increasing factor, $N$ is number of steps required for heating. Increasing $N$ allows to approach the quasistatic limit.~\cite{Peng2016,Begun2019}.

\section{Studied proteins}

Two proteins containing complex prime knots were selected for the analysis: ubiquitin carboxyl-terminal hydrolase L1 with a \(5_2\) knot and DL-2-haloacid dehalogenase with a \(6_1\) knot in the structure.

\paragraph{UCH-L1 (PDB ID: 3IRT).}
The structure 3IRT corresponds to the I93M variant of human ubiquitin carboxyl-terminal hydrolase L1, a cysteine deubiquitinating enzyme. The structure was determined by X-ray diffraction at a resolution of \(2.80\,\text{\AA}\). The protein sequence contains  223 residues resolved in chain A~\cite{Boudreaux2010,PDB3IRT}. According to KnotProt, chain A contains a \(5_2\) knot with a knotted core extending from residues 5 to 220~\cite{DabrowskiTumanski2019,KnotProt3IRT}. Thus, the knot involves almost the entire resolved polypeptide chain, making 3IRT a suitable model for studying the unfolding of a \(5_2\)-knotted protein.

The soliton model of this protein contains 24 solitons and achieve accuracy of 0.69 \AA\:  RMSD from experimental model.

\paragraph{DL-2-haloacid dehalogenase (PDB ID: 4N2X).}
The structure 4N2X represents DL-2-haloacid dehalogenase from \textit{Methylobacterium} sp.\ CPA1. This enzyme catalyzes the dehalogenation of both enantiomers of 2-haloacids. The structure was determined by X-ray diffraction at a resolution of \(1.70\,\text{\AA}\). Each subunit contains 301 residues, with 298 residues resolved in chain A, while the biological assembly is a homohexamer~\cite{Siwek2013,PDB4N2X}. Chain A contains a \(6_1\) knot whose core extends from residues 63 to 278~\cite{DabrowskiTumanski2019,KnotProt4N2X}. These embedded topological motifs make 4N2X particularly useful for investigating whether the \(6_1\) knot disappears directly or passes through simpler intermediate knot types during unfolding.

For the subsequent analysis, chain A of each structure was treated as an individual open polymer chain, and changes in its topology were monitored along the unfolding trajectory. The soliton model of 4N2X contains 26 solitons and achieve accuracy of 0.67 \AA\: RMSD from experimental model.

\begin{figure}[ht!]
    \centering
    \begin{subfigure}{0.98\textwidth}
        \centering
        \includegraphics[width=\textwidth]{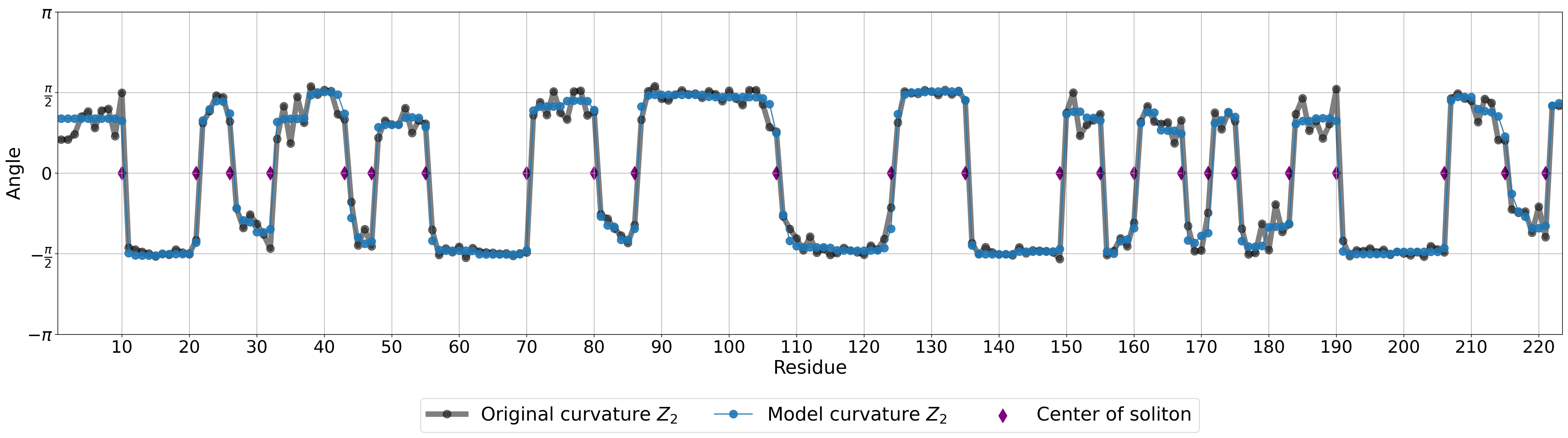}
        \caption{Curvature \(\kappa_i\) of 3IRT.}
        \label{fig:3irt_kappa}
    \end{subfigure}

    \vspace{0.4em}
    \begin{subfigure}{0.98\textwidth}
        \centering
        \includegraphics[width=\textwidth]{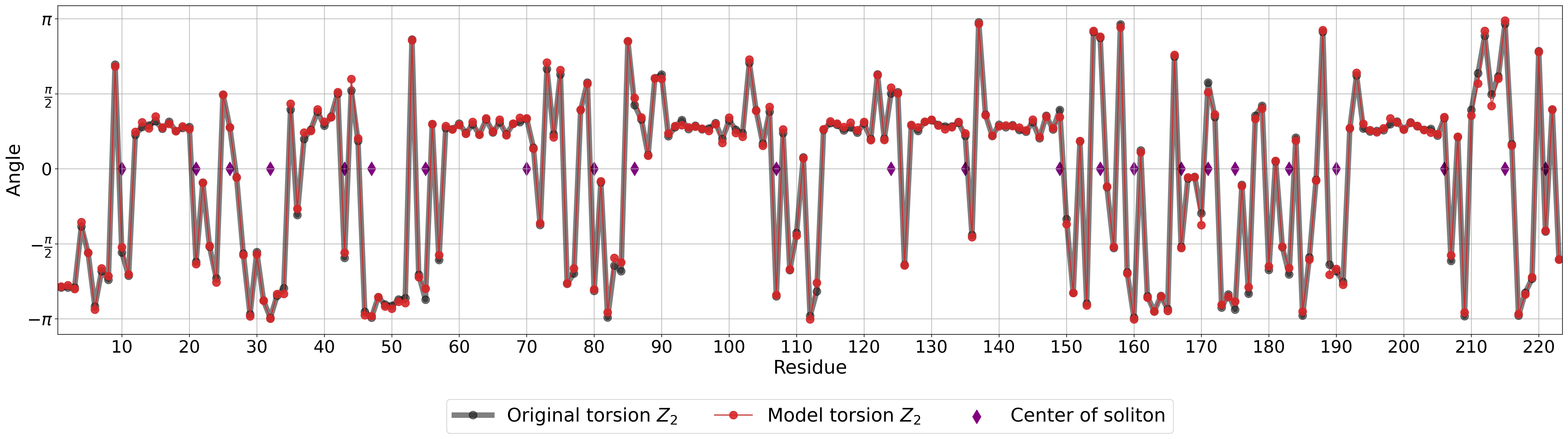}
        \caption{Torsion \(\tau_i\) of 3IRT.}
        \label{fig:3irt_tau}
    \end{subfigure}

    \vspace{0.4em}
    \begin{subfigure}{0.98\textwidth}
        \centering
        \includegraphics[width=\textwidth]{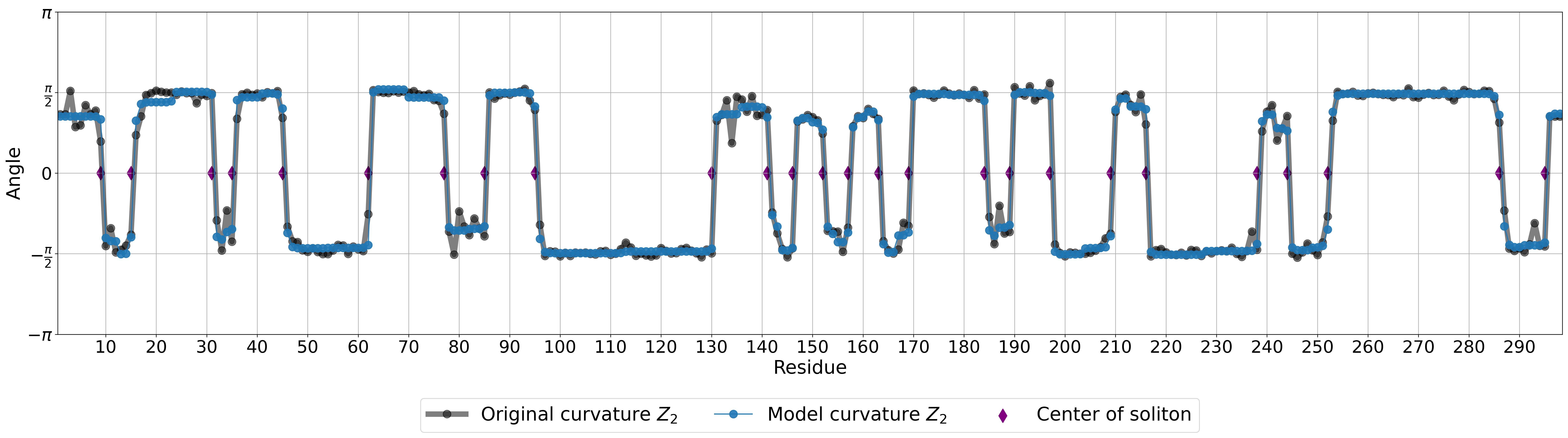}
        \caption{Curvature \(\kappa_i\) of 4N2X.}
        \label{fig:4n2x_kappa}
    \end{subfigure}

    \vspace{0.4em}
    \begin{subfigure}{0.98\textwidth}
        \centering
        \includegraphics[width=\textwidth]{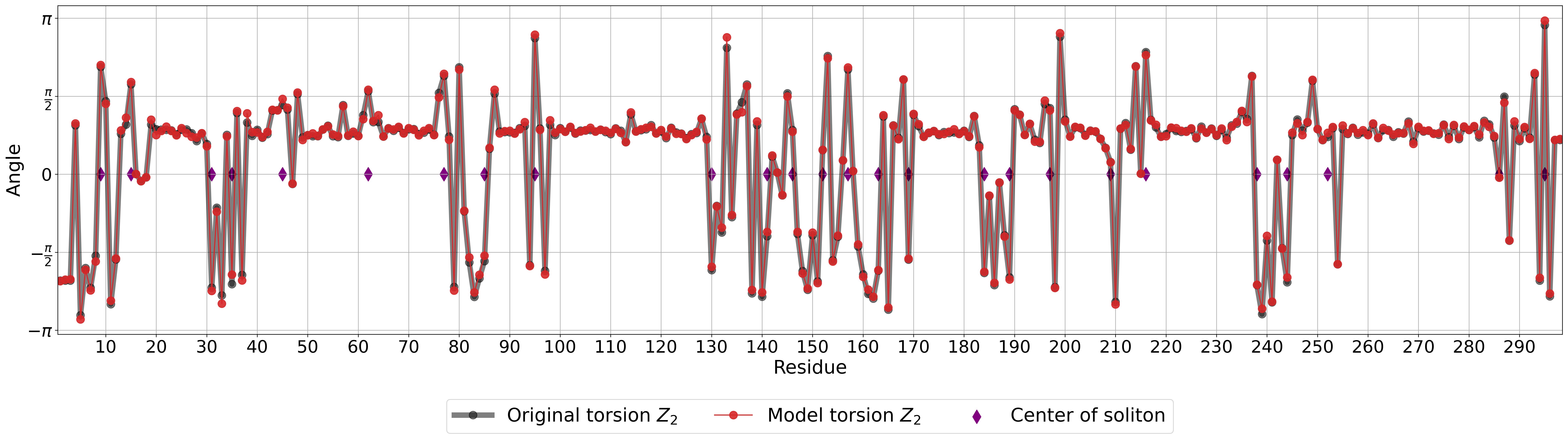}
        \caption{Torsion \(\tau_i\) of 4N2X.}
        \label{fig:4n2x_tau}
    \end{subfigure}

    \caption{Comparison of the local backbone geometry obtained from the experimental PDB structures and from the corresponding soliton models. The experimental and model values of curvature and torsion are shown for 3IRT and 4N2X.}
    \label{fig:model_angles}
\end{figure}

The agreement between the experimental backbone geometry and the corresponding soliton models is shown in Fig.~\ref{fig:model_angles}. For both proteins, the model reproduces the main features of the curvature \(\kappa_i\) and torsion \(\tau_i\) profiles along the chain.

\begin{figure}[ht!]
    \centering
    \begin{subfigure}[t]{0.48\textwidth}
        \centering
        \includegraphics[width=\textwidth]{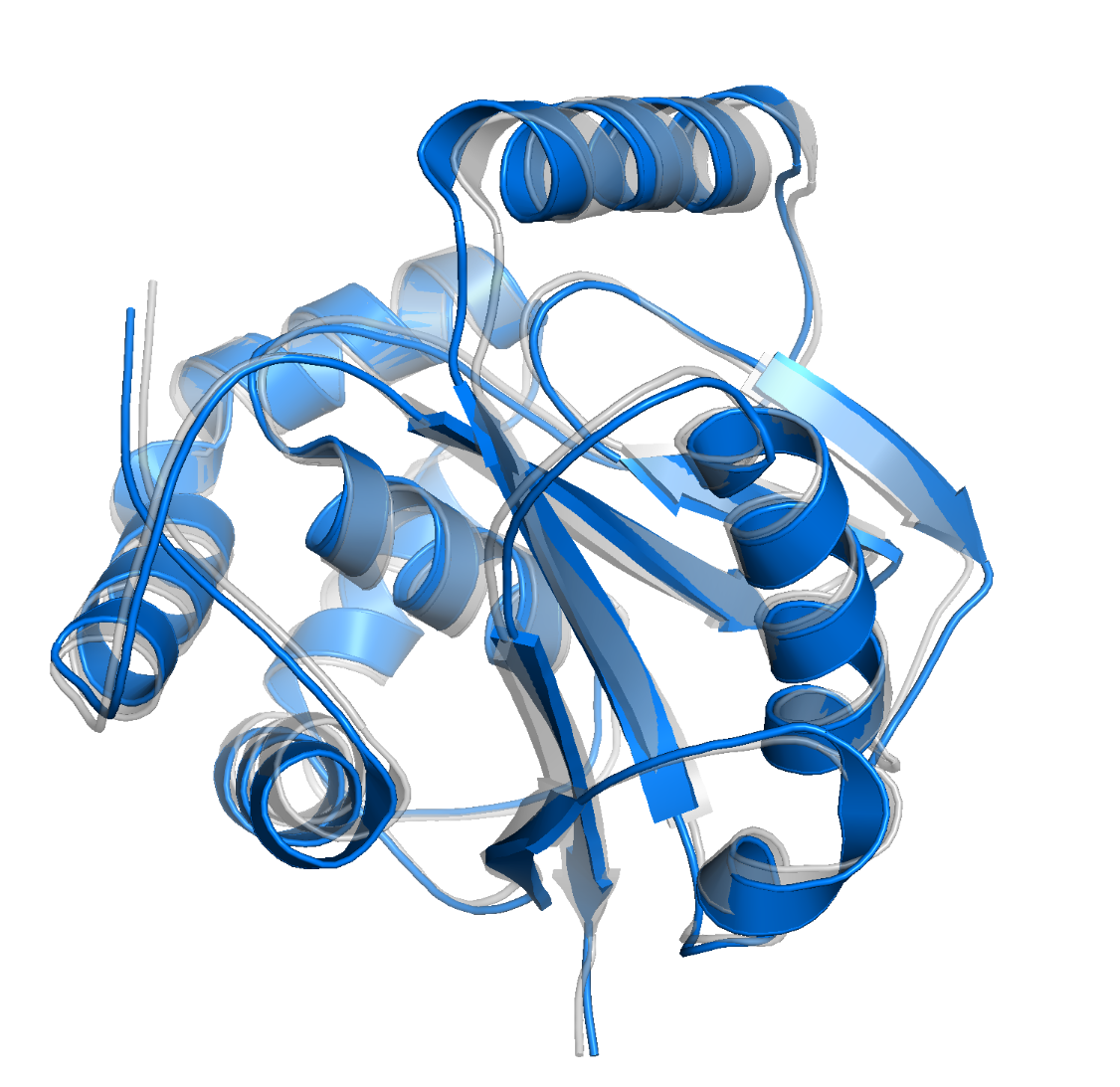}
        \caption{3IRT, RMSD \(0.69\,\text{\AA}\).}
        \label{fig:3irt_overlay}
    \end{subfigure}
    \hfill
    \begin{subfigure}[t]{0.48\textwidth}
        \centering
        \includegraphics[width=\textwidth]{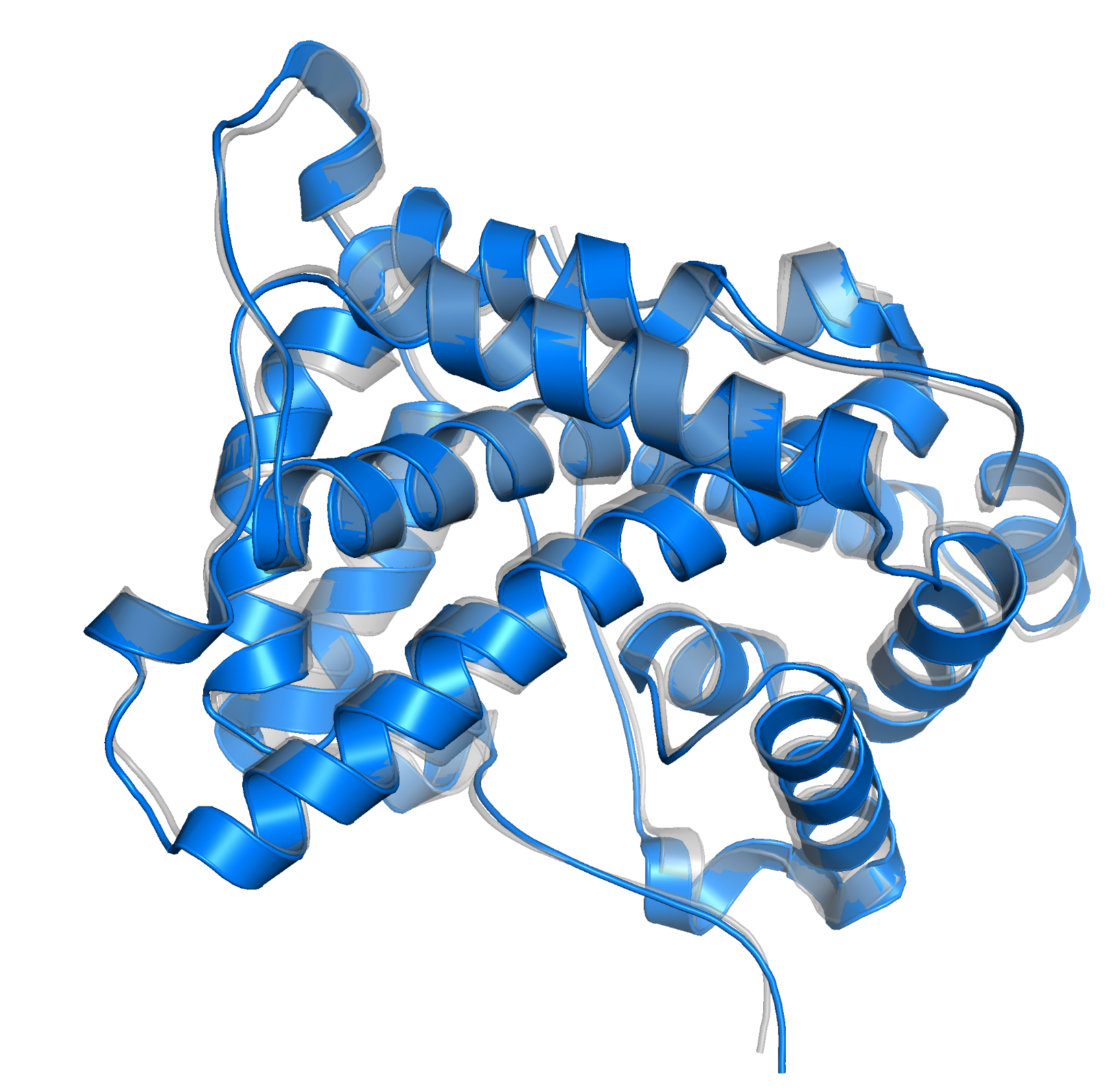}
        \caption{4N2X, RMSD \(0.67\,\text{\AA}\).}
        \label{fig:4n2x_overlay}
    \end{subfigure}
    \caption{Superposition of the experimental PDB structures and the corresponding three-dimensional structures reconstructed from the soliton model.}
    \label{fig:model_overlay}
\end{figure}

The agreement between the experimental and reconstructed structures is also observed at the level of the three-dimensional backbone conformation. Figure~\ref{fig:model_overlay} shows the superposition of the PDB structures and the structures reconstructed from the soliton model.

\section{Results. Topological unfolding of 3IRT and 4N2X}

\begin{figure}[ht]
    \centering
    \includegraphics[width=0.9\textwidth]{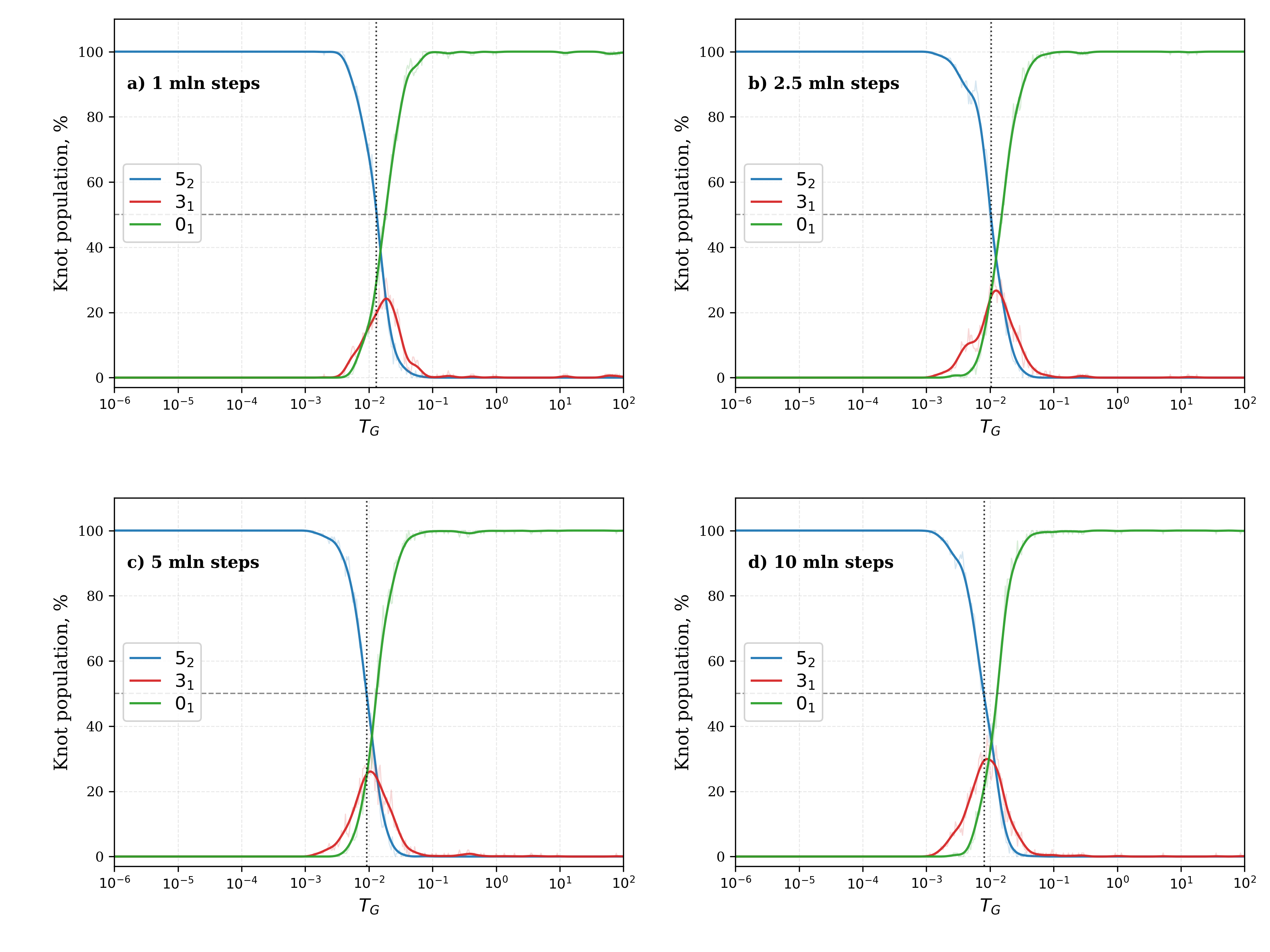}
    \caption{Temperature dependence of the populations of different topological states observed during the unfolding of 3IRT for different numbers of Monte Carlo steps \(N\). The native state corresponds to the \(5_2\) knot, while \(0_1\) denotes the topologically trivial state.}
    \label{fig:3irt_topology}
\end{figure}

\begin{figure}[ht]
    \centering
    \includegraphics[width=0.9\textwidth]{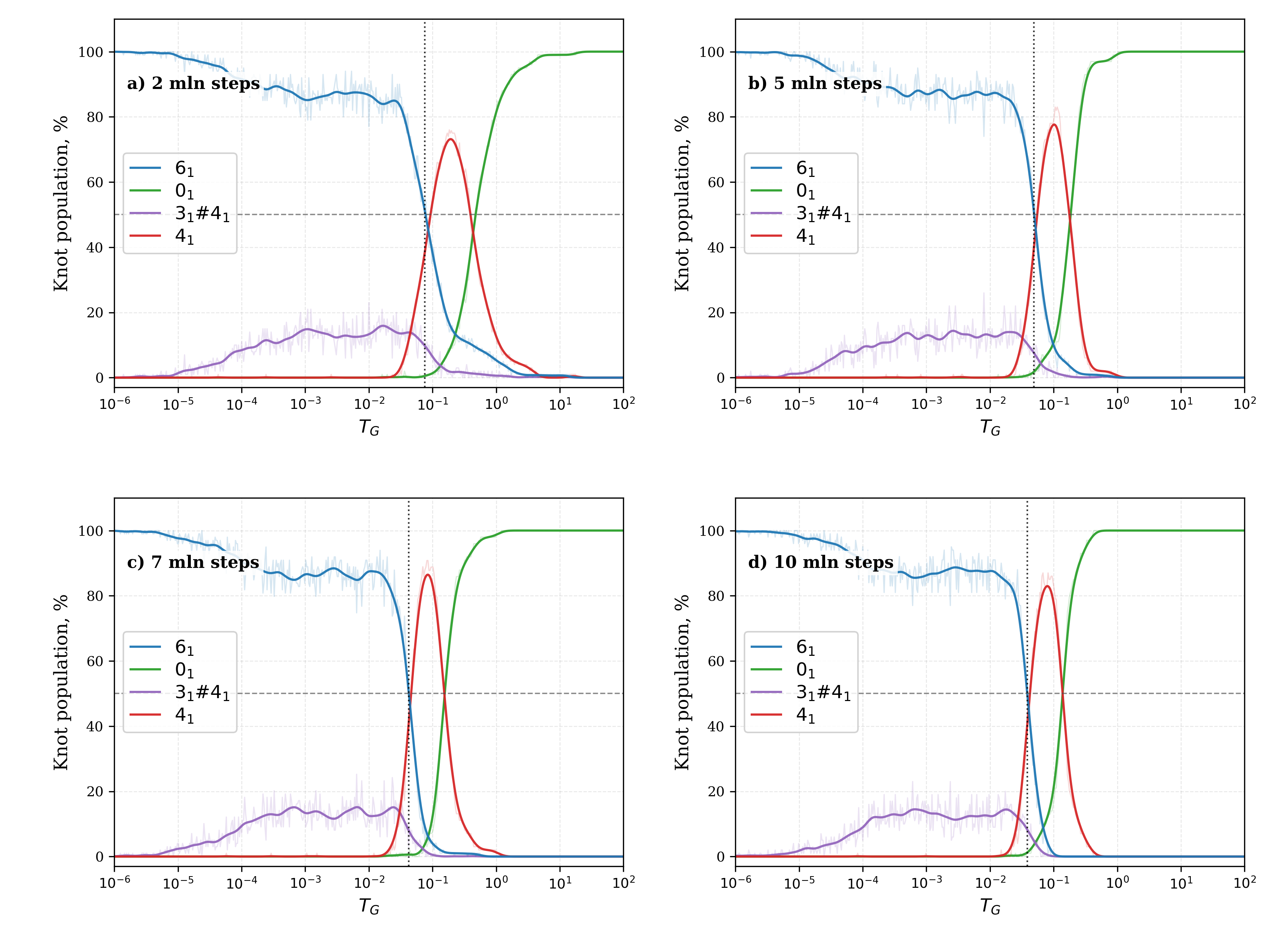}
    \caption{Temperature dependence of the populations of different topological states observed during the unfolding of 4N2X for different numbers of Monte Carlo steps \(N\). In addition to the native \(6_1\) knot and the trivial \(0_1\) state, intermediate \(3_1\mathbin{\#}4_1\) and \(4_1\) topologies are observed.}
    \label{fig:4n2x_topology}
\end{figure}

Unfolding simulations were performed for both proteins using the same heating protocol. The duration of the heating stage was varied from \(1\times10^{6}\) to \(2\times10^{7}\) simulation steps. At each stage of the unfolding trajectory, the topology of the protein backbone was determined using the \texttt{Knot\_pull} algorithm~\cite{Jarmolinska2020}. The resulting distributions of the detected knot types as functions of temperature are shown in Figs.~\ref{fig:3irt_topology} and~\ref{fig:4n2x_topology}.

For both proteins, increasing the number of simulation steps shifts the topological transition towards lower temperatures. Since the same temperature interval is covered in all simulations, a larger value of \(N\) corresponds to a slower heating process. 

\begin{figure}[ht]
    \centering
    \begin{tabular}{c c}
     \includegraphics[width=0.45\textwidth]{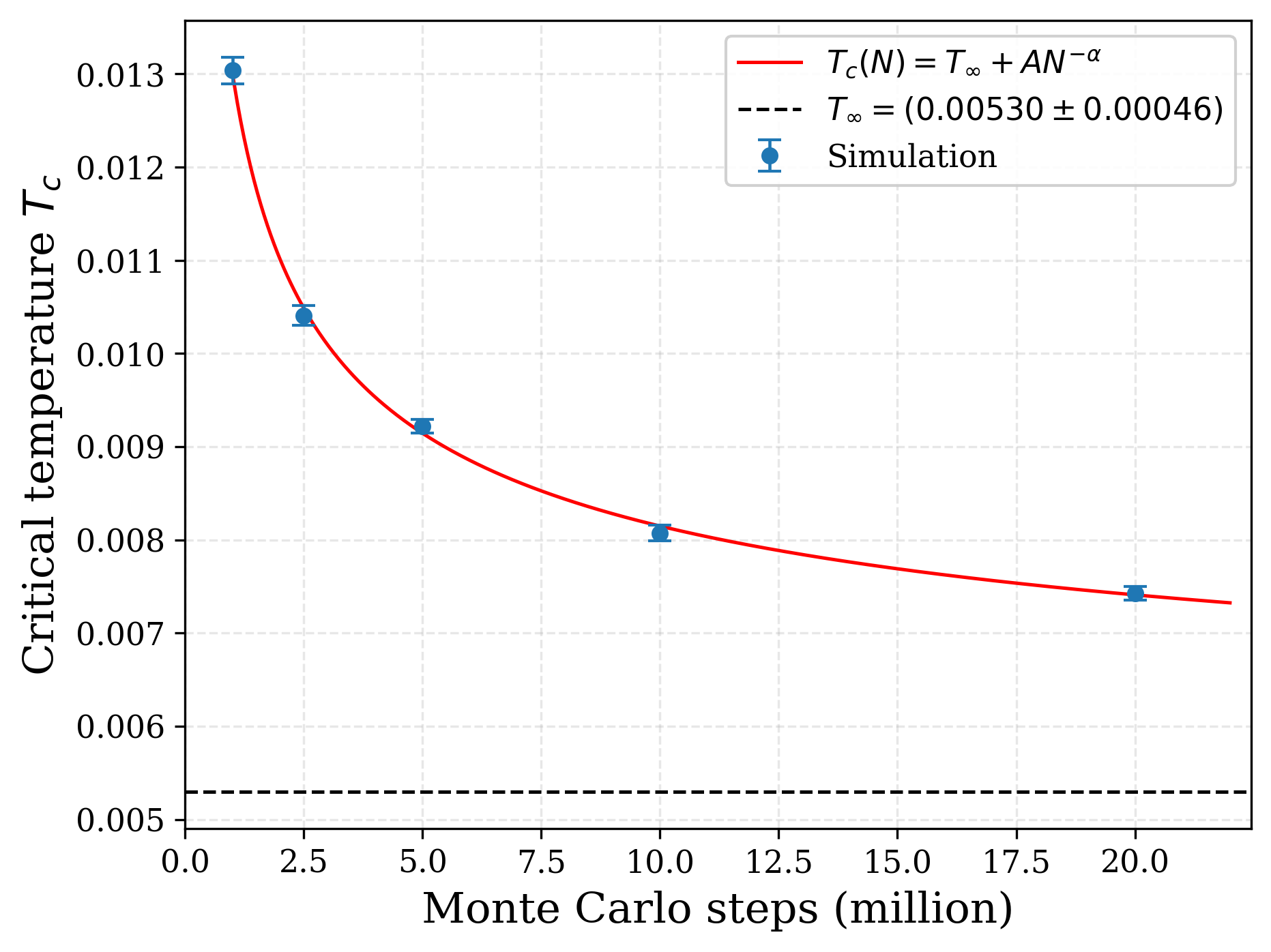}    &    \includegraphics[width=0.45\textwidth]{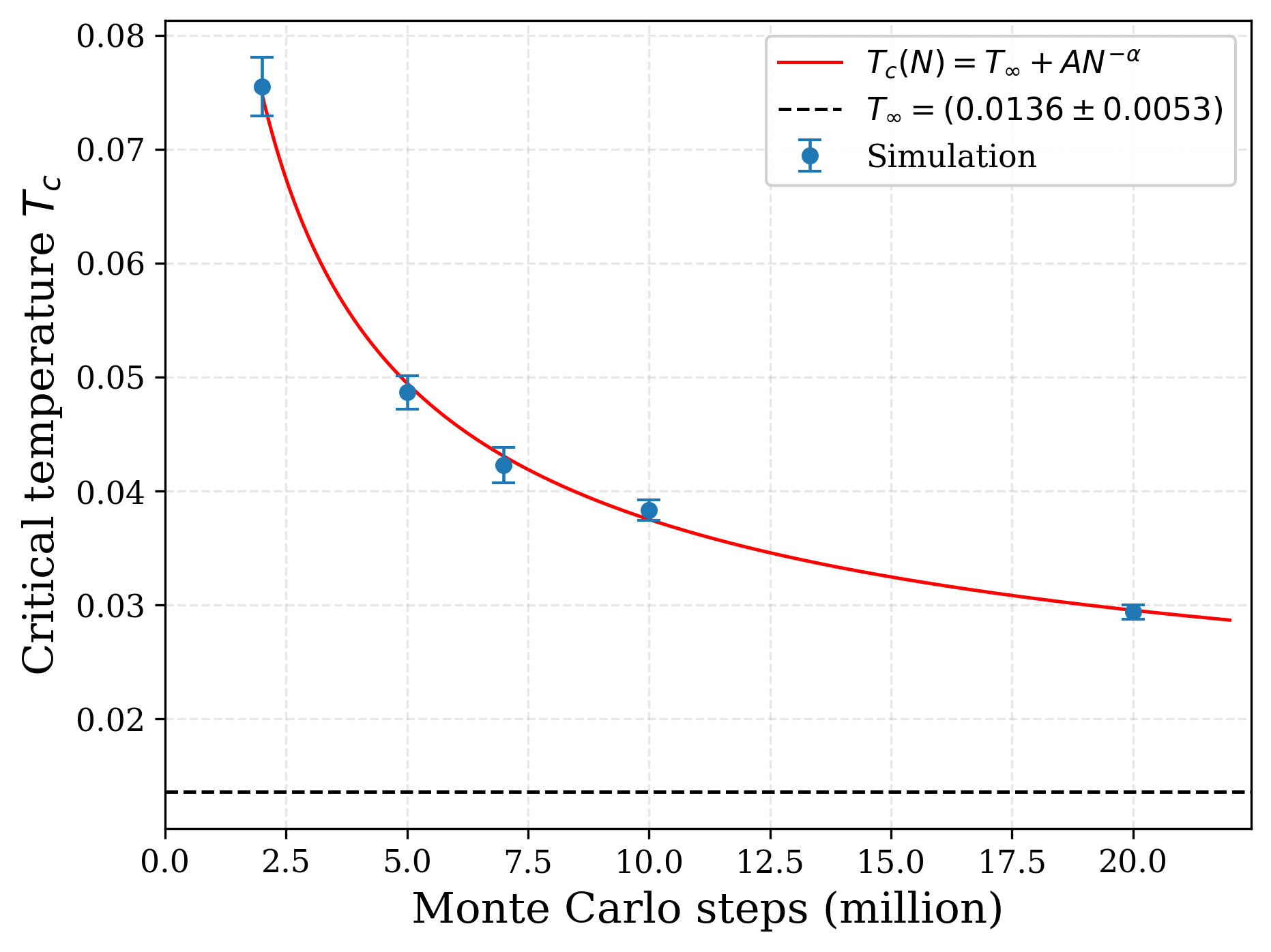}
    \end{tabular}
    
    \caption{Dependence of the critical temperature \(T_{\mathrm{crit}}\) on the number of Monte Carlo steps \(N\) for 3IRT and 4N2X. The points correspond to the critical temperatures extracted from the topology distributions, while the curves represent the corresponding fits. The extrapolated value \(T_{\infty}\) corresponds to the limit \(N\rightarrow\infty\), i.e., infinitely slow heating.}
    \label{fig:transition_temperature}
\end{figure}

To quantify this effect, a critical temperature \(T_{\mathrm{crit}}\) was determined for each value of \(N\) from the temperature dependence of the native-knot population. The obtained values were then analyzed as a function of the number of Monte Carlo steps. The corresponding dependences for 3IRT and 4N2X, together with the fitted curves, are shown in Fig.~\ref{fig:transition_temperature}.

The fitted dependences approach finite limiting values as \(N\) increases. We denote the extrapolated critical temperature in the limit of infinitely slow heating as
\begin{equation}
    T_{\infty}
    =
    \lim_{N\rightarrow\infty}T_{\mathrm{crit}}(N).
\end{equation}
For the two proteins, the obtained limiting temperatures are
\begin{equation}
    T_{\infty}^{3IRT} = (530 \pm 46)10^{-5},
\end{equation}
and
\begin{equation}
    T_{\infty}^{4N2X} = (136 \pm 53)10^{-4}.
\end{equation}

Here, temperature is expressed in the algorithmic Glauber scale used in the Monte Carlo simulations. A direct correspondence between the Glauber temperature and the physical temperature measured in degrees Celsius was not established in the present work. Nevertheless, the model and its temperature scale allow us to qualitatively characterize the behavior of the proteins and compare their topological response to heating.

While \(T_{\infty}\) characterizes the limiting temperature of the topological transition under progressively slower heating, it does not describe the mechanism by which the knot disappears. We therefore analyzed the sequence of topological states along the unfolding trajectories for both proteins.

The topological unfolding pathway of 3IRT is relatively simple. In most trajectories, the native \(5_2\) knot transforms directly into the topologically trivial state \(0_1\),
\begin{equation}
    5_2 \longrightarrow 0_1.
\end{equation}
Only a small number of trajectories contain an intermediate trefoil state,
\begin{equation}
    5_2 \longrightarrow 3_1 \longrightarrow 0_1.
\end{equation}

The topological unfolding of 4N2X is more complex. Configurations with the composite topology \(3_1\mathbin{\#}4_1\) were observed in all trajectories, and transitions between the native \(6_1\) state and the composite state occurred repeatedly during the early stages of unfolding. The overall topological evolution can be summarized as
\begin{equation}
    6_1
    \rightleftarrows
    3_1\mathbin{\#}4_1
    \longrightarrow
    4_1
    \longrightarrow
    0_1.
\end{equation}
The transition to the \(4_1\) state does not necessarily occur directly from the composite state.

\section{Discussion}

For both proteins, increasing the duration of the heating protocol shifts the loss of the native knot towards lower Glauber temperatures. This indicates that the observed transition depends on the time available for the chain to leave the native topological state. Since the Glauber temperature is an algorithmic quantity~\cite{Peng2016,Begun2019}, these values should be interpreted only within the present model and not as experimental melting temperatures.

For 3IRT, the dominant topological pathway is the direct transition \(5_2\longrightarrow0_1\). A transient \(3_1\) knot appears only in a small fraction of trajectories. This is consistent with previous simulations in which the resulting topology depended on the direction of unthreading~\cite{Zhao2018}. Single-molecule experiments on UCH-L1 also showed that different pulling geometries can produce unfolded chains with \(5_2\), \(3_1\), or \(0_1\) topology~\cite{UCHL1SingleMolecule}. The experimental perturbation was mechanical rather than thermal, so the comparison is qualitative.

The predominance of the direct topological transition does not imply two-state conformational unfolding. Experimental studies of UCH-L1 have identified several conformational intermediates~\cite{UCHL1Folding,ZhangJackson2016,UCHL1SingleMolecule}. Such states can have the same knot type and are therefore not distinguished by the present topological analysis.

The 3IRT structure used here corresponds to the I93M variant of UCH-L1. This mutation is known to destabilize the protein and increase its intrinsic unfolding rate relative to the wild type~\cite{UCHL1Folding,UCHL1Proteolysis}. Therefore, the transition temperatures obtained for 3IRT should not be directly transferred to wild-type UCH-L1.

The behavior of 4N2X is different. The \(3_1\mathbin{\#}4_1\) topology is observed in all trajectories, with repeated transitions between this state and the native \(6_1\) knot. The composite state is therefore not an obligatory step of the final unfolding transition, but it is easily accessible from the native topology. The subsequent \(4_1\) state is more persistent and precedes complete unknotting.

A \(4_1\) intermediate has also been reported in coarse-grained simulations of another \(6_1\)-knotted dehalogenase, DehI~\cite{DehISimulation}. Chemical-denaturation experiments on DehI revealed conformational intermediates as well~\cite{DehIFolding}, although their knot types were not determined. The protein studied in those works differs from 4N2X, so this comparison should be regarded as topological rather than protein-specific.

To our knowledge, the composite \(3_1\mathbin{\#}4_1\) state has not previously been reported during the unfolding of a \(6_1\)-knotted dehalogenase. Its repeated appearance in independent trajectories suggests that it is a characteristic feature of the topological dynamics in the present model.

The present analysis is limited to individual protein chains. In particular, 4N2X forms an oligomeric biological assembly, and inter-subunit interactions are not included. In addition, very short-lived topological states can be missed because the knot type is evaluated only at discrete points along the trajectory.

\section{Conclusions}

The \(5_2\)-knotted protein 3IRT and the \(6_1\)-knotted protein 4N2X show qualitatively different topological unfolding behavior. In 3IRT, the native knot is usually removed directly, while a \(3_1\) intermediate is rare. In 4N2X, the native \(6_1\) state repeatedly interconverts with the composite \(3_1\mathbin{\#}4_1\) topology before the system reaches the simpler \(4_1\) knot and finally the unknotted state.

For both proteins, the transition temperature decreases as the heating duration is increased, allowing the limiting value \(T_{\infty}\) to be estimated. Within the same computational model, 4N2X therefore shows a more complex topological unfolding dynamics than 3IRT.

\section*{Author Contributions}

A.B. conceived the study, selected the protein structures, developed the simulation code, analyzed the results, prepared the figures, and wrote the manuscript. S.L. constructed the soliton models of the protein structures. A.B. and N.G. described the theoretical model. A.Z. performed the simulations, prepared the simulation data, and produced the corresponding plots. I.B. performed the simulations. All authors discussed the results and reviewed the manuscript.

\section*{Acknowlegements}

The authors are supported  by the state assignment of the Ministry of Science and Higher Education of Russia (Project No. FZNS-2024-0002). The work of S.L. was supported by a grant from the Ministry of Science and Higher Education of the Russian Federation under the FEFU Priority 2030 Program (topic No. ASP-25-02-1.03-0022).

\bibliographystyle{unsrtnat}
\bibliography{references}

\end{document}